\documentclass[8pt,a4paper]{article}
\usepackage[utf8]{inputenc}
\usepackage{amsmath}
\usepackage{amsfonts}
\usepackage{amssymb}
\usepackage{graphicx}
\usepackage{xcolor}
\usepackage{url}
\usepackage{braket}
\usepackage{tikz}
\usepackage[left=3cm, right=3cm, top=2cm, bottom=2cm]{geometry}
\usepackage{rotating}
\usepackage{lscape}
\usepackage{hyperref}
\usepackage{lscape}
\usepackage{wrapfig}
\usepackage{float}
\usepackage{pdfpages}
\newcommand{\rl}[1]{\left(#1\right)}

\allowdisplaybreaks

\def\be{\begin{equation}}
\def\ee{\end{equation}}
\def\ba{\begin{array}{l}}
	\def\ea{\end{array}}
\def\bea{\begin{eqnarray}}
\def\eea{\end{eqnarray}}
\def\beas{\begin{eqnarray*}}
	\def\eeas{\end{eqnarray*}}

\def\ket#1{| #1 \rangle}
\def\bra#1{ \langle #1 |}

\begin{document}
	\baselineskip 24pt
	
	\begin{center}
		
		{\large \textbf{The Anyonic Quantum Carnot Engine}\par}
		
	\end{center}
	
	\vskip .5cm
	\medskip
	
	\vspace*{4.0ex}
	
	\baselineskip=18pt
\begin{center}
\large
H S Mani$^{(a)}$\footnote {hsmani@cmi.ac.in},
N Ramadas$^{(b)}$\footnote{ramadas.phy@gmail.com},
and V V Sreedhar$^{(a)}$\footnote{sreedhar@cmi.ac.in}

\end{center}
        \vspace*{4.0ex}

        \centerline{\it $^{(a)}$ Chennai Mathematical Institute,  SIPCOT IT Park, Siruseri, Chennai, 603103 India}

        \centerline{\it $^{(b)}$ The Institute of Mathematical Sciences, CIT Campus, Tharamani, Chennai, 600036 India}

        \vspace*{1.0ex}

        \vspace*{1.0ex}

        \vspace*{5.0ex}

        \begin{abstract}
Combining two disparate lines of thought like thermodynamics and quantum 
mechanics yields surprising results. The resulting idea of quantum 
thermodynamic engines holds promise for harvesting novel sources of energy 
of purely quantum origin, like quantum statistics, to extract mechanical 
work from macroscopic quantum systems like Bose-Einstein condensates. It 
also allows one to explore thermodynamic thinking to `small' systems consisting
only a few particles in quantum theory. In an earlier paper we had 
studied the quantum Otto engine using a small number of one-dimensional 
anyons as the medium to drive the thermodynamic engine. In this sequel,
we study the gold-standard of thermodynamic engines, namely the Carnot 
engine, using two-dimensional anyons as the medium. An explicit expression
is derived for the efficiency of the anyonic quantum Carnot engine.

	\end{abstract}


\section{Introduction}

In a recent paper \cite{maniQuantumThermodynamicsSmall2024}, the authors started exploring the role of 
quantum statistics in quantum thermodynamics, a subject that connects the 
seemingly disparate areas of thermodynamics and quantum mechanics. This line 
of study is interesting because it raises the intriguing possibility of 
harnessing non-classical, non-thermal sources of energy of quantum systems to 
do mechanical work \cite{binderThermodynamicsQuantumRegime2018,myersQuantumThermodynamicDevices2022}%,kochQuantumEngineBEC2023}
. It also raises fundamental questions 
regarding the limits of applicability of thermodynamic ideas to small 
systems \cite{feshbachSmallSystemsWhen1988,hillThermodynamicsSmallSystems2013} {\it i.e.,} systems with just a few 
particles, say, as opposed to macroscopic systems, for a phenomenological 
description of which thermodynamics was originally invented \cite{callenThermodynamicsIntroductionThermostatistics1985}. 
 
In \cite{maniQuantumThermodynamicsSmall2024}, we studied a toy model of one-dimensional anyons to 
drive a quantum Otto engine, {\it aka} the Pauli engine, back and forth 
between the bosonic and fermionic limits. In doing so, we mimicked the 
experimental results of \cite{kochQuantumEngineBEC2023} in which a many-body quantum engine 
with harmonically trapped $~^6$Li atoms close to a magnetic Feshbach 
resonance was used to drive the quantum gas back and forth between a 
Bose-Einstein condensate and a unitary Fermi gas. The experiment exploits
the underlying phenomenon of BEC-BCS crossover \cite{leggettQuantumLiquidsBose2006}, to affect 
the change in statistics by tuning a magnetic field. 

In this paper, we use two-dimensional anyons in a harmonic trap as the working 
medium for quantum thermodynamics. Our goal is to explore the effect of 
quantum statistics in small systems on thermodynamic engines. Towards this 
end we construct the anyonic quantum Carnot engine -- the gold standard of
thermodynamic engines -- and compute its efficiency. 

The paper is organised as follows: In the next section, we give a lightning 
review of two anyons in a harmonic trap. In section 3, we describe a quantum 
version of the Carnot engine. In section 4, we construct the anyonic quantum 
Carnot engine and compute its efficiency. In section 5, we summarise and 
conclude. 
 
\section{Anyons}
As is well-known, the symmetrization postulate imposes restrictions on the 
nature of wave functions describing indistinguishable particles in quantum 
mechanics: the wave functions are either symmetric or antisymmetric, 
corresponding to Bose-Einstein statistics, or Fermi-Dirac statistics 
respectively.  In a classic paper, Leinaas and Myrheim \cite{leinaasTheoryIdenticalParticles1977} 
traced the origin of the symmetrization postulate to the topology of the
configuration space of indistinguishable particles. A spin-off of this 
observation is that, the statistics of indistinguishable particles is 
described by a representation of the fundamental group of the configuration 
space. In two dimensions this is the braid group whose lowest dimensional 
representation is given by a phase factor, as opposed to the permutation 
group whose lowest dimensional representations {\it viz.} the identity 
representation and the alternating representation yield the aforementioned 
bosonic and fermionic cases in higher dimensions. As a result, the wave 
function of indistinguishable particles in two dimensions picks up a 
non-trivial phase factor when an exchange of particles is performed. Since 
bosons and fermions correspond to special cases in which the phase factor 
collapses to $\pm 1$, such particles are called anyons. 

\subsection{Two anyons in a harmonic oscillator potential}
\label{sec:Two anyons in a harmonic trap}
In this section we summarize the results of \cite{myrheimAnyons1999}. The Hamiltonian 
for two identical partices in a two-dimensional harmonic trap is  
\begin{align}
\begin{aligned}
H = \frac{1}{2m} \rl{\pmb{p}_1^2 +\pmb{p}_2^2}+ \frac{1}{2} m \omega^2 
\rl{\pmb{x}_1^2+\pmb{x}_2^2}
\end{aligned}
\end{align}
where $ \pmb{x}_1 $ and $ \pmb{x}_2 $ are the positions, $ \pmb{p}_1 $ and 
$ \pmb{p}_2 $ are the momenta, $ \omega $ is the frequency, and $ m $ is the 
mass of the particles. Anyonic statistics can be realised by the simple 
exigency of requiring the two-particle wave function to pick up a phase 
factor under the exchange of the particle positions, 
\begin{align}
\begin{aligned}
\psi\rl{\pmb{x}_2, \pmb{x}_1} = e^{i\pi \nu}  \psi\rl{\pmb{x}_1, \pmb{x}_2},
\end{aligned}
\end{align}
where $ 0 \leq \nu \leq 1 $. Note that $ \nu=0 $ represents bosons, while 
$ \nu=1 $ represents fermions. It is to be expected that the energy spectrum
will depend on the statistics parameter $\nu$.   

Defining the center of mass and relative coordinates for positions and momenta,
\begin{align}
\begin{aligned}
\pmb{X} = \frac{1}{2} \rl{ \pmb{x}_1+\pmb{x}_2}, ~~~ \pmb{x} = 
\pmb{x}_1-\pmb{x}_2 \\
\pmb{P} =  \rl{ \pmb{p}_1+\pmb{p}_2}, ~~~ \pmb{p} =  \frac{1}{2} 
\rl{\pmb{p}_1-\pmb{p}_2 },
\end{aligned}
\end{align}
the Hamiltonian can be rewritten as
\begin{align}
\begin{aligned}
H = \frac{\pmb{P}^2}{4m} +\frac{\pmb{p}^2}{m} + m \omega^2 \pmb{X}^2 +   
\frac{1}{4}m \omega^2 \pmb{x}^2.
\end{aligned}
\end{align}
In these coordinates, the anyonic statistics condition reads as follows:
\begin{align}
\begin{aligned}
\psi\rl{\pmb{X},-\pmb{x}} =  e^{i\pi \nu} \psi\rl{\pmb{X},\pmb{x}}.
\end{aligned}
\end{align}
This condition is singular at $ \pmb{x} =0 $ whenever $ \nu $ is not an 
even integer. This forces the  wave function to be singular at $ \pmb{x} =0 $,
in accordance with the Pauli exclusion principle. 

Let us consider the relative motion of the system and let $ r $ and $ \phi $ 
be the relative polar coordinates. The wave function must have the following 
form as $ r\to 0 $
\begin{align}
\begin{aligned}
\psi_{rel} (r,\phi) = r^\mu e^{i \ell \phi}
\end{aligned}
\end{align}
with $\mu > 0$ and $ \ell = \nu +2 k $, for some integer $ k $. From the 
energy value condition, we get $ \mu^2-\ell^2=0 $, to leading order in $ r $. 
We choose the solution $ \mu = |\ell| $, to make $ \psi $ finite in the limit 
$ r\to0 $. Therefore, there are two classes of energy eigenstates: class (I) 
having $ \mu =\nu,\nu+2,\nu+4,...  $ and class (II) having 
$ \mu =2-\nu,4-\nu,6-\nu,... $.

We introduce complex coordinates 
\begin{align*}
\begin{aligned}
z_1 = \frac{1}{\sqrt{\lambda}} \rl{x_1 +i y_1},~~~ z_2 = \frac{1}
{\sqrt{\lambda}} \rl{x_2 +i y_2},~~~\text{where}~~\lambda = \sqrt{\frac{\hbar}
{m \omega}}
\end{aligned}
\end{align*}
Further, if $ Z = \frac{1}{2} \rl{z_1 +z_2} $ and $ z=z_1-z_2 $, the
Hamiltonian takes the form
\begin{align}
\begin{aligned}
H &= \hbar \omega \rl{ -2 \frac{\partial^2}{\partial z_1 \partial z_1^*} -2 
\frac{\partial^2}{\partial z_2 \partial z_2^*} +\frac{|z_1|^2}{2} +
\frac{|z_2|^2}{2}} \\
&= \hbar \omega \rl{ 2 \frac{\partial^2}{\partial Z \partial Z^*} 
-4 \frac{\partial^2}{\partial z \partial z^*} +|Z|^2 +\frac{|z|^2}{4}}
\end{aligned}
\end{align}
We define the creation and annihilation operators as follows: 
\begin{align}
\begin{aligned}
a &= \frac{1}{\sqrt{2}}\frac{\partial}{\partial Z} + \frac{Z^*}{\sqrt{2}}, 
\qquad a^\dagger =- \frac{1}{\sqrt{2}}\frac{\partial}{\partial Z^*} + 
\frac{Z}{\sqrt{2} } \\
b &= \frac{1}{\sqrt{2}}\frac{\partial}{\partial Z^*} + \frac{Z}{\sqrt{2}}, 
\qquad b^\dagger = -\frac{1}{\sqrt{2}}\frac{\partial}{\partial Z} + 
\frac{Z^*}{\sqrt{2} } \\
c &= \sqrt{2}\frac{\partial}{\partial z} + \frac{z^*}{2\sqrt{2}}, \qquad 
c^\dagger =-\sqrt{2}\frac{\partial}{\partial z^*} + \frac{z}{2\sqrt{2}}\\
d &= \sqrt{2}\frac{\partial}{\partial z^*} + \frac{z}{2\sqrt{2}}, \qquad 
d^\dagger =-\sqrt{2}\frac{\partial}{\partial z} + \frac{z^*}{2\sqrt{2}}\\
\end{aligned}
\end{align}
These operators satisfy the commutation relations
\begin{align}
\begin{aligned}
\left[a,a^\dagger\right] = [b,b^\dagger] =[c,c^\dagger] =[d,d^\dagger] =1,
\end{aligned}
\end{align}
with all other commutators equal to zero.
The Hamiltonian in terms of these operators is given by
\begin{align}
\begin{aligned}
H &= \hbar \omega (a^\dagger a +b^\dagger b + c^\dagger c +d^\dagger d+2) \\
\end{aligned}
\end{align}
Using the above formulation, we can find the energy eigenstates corresponding 
to class(I) and class(II) described earlier \begin{align}
\begin{aligned} 
\psi^{(I)}_{j,k,l,m} & = (a^\dagger)^j (b^\dagger)^k (c^\dagger d^\dagger)^{l} 
(d^\dagger)^{2m} \psi_0^{(I)}, \\
\psi^{(II)}_{j,k,l,m} & = (a^\dagger)^j (b^\dagger)^k (c^\dagger d^\dagger)^{l}
 (d^\dagger)^{2m} \psi_0^{(II)}, \\
\end{aligned}
\end{align}
where 
\begin{align}
\begin{aligned}
\psi_0^{(I)} = z^\nu e^{- \rl{|Z|^2+\frac{|z|^2}{4} }},~~~~ \psi_0^{(II)} = 
(z^*)^{2-\nu} e^{- \rl{|Z|^2+\frac{|z|^2}{4} }},
\end{aligned}
\end{align}
and $ j,k,l,m=0,1,2,... $ Here $ j,k $ describe center of mass excitations 
while $ l,m $ describe excitations of the relative degrees of freedom.
The corresponding energy eigenvalues are
\begin{align}
\begin{aligned}
E^{(I)}_{j,k,l,m} & = \rl{2+\nu+j+k+2l+2m} \hbar \omega\\
E^{(II)}_{j,k,l,m} & = \rl{4-\nu+j+k+2l+2m} \hbar \omega
\end{aligned}
\end{align}
\section{Quantum Carnot engine}
The classical Carnot cycle consists of four process: 1) Isothermal expansion 
of the classical gas at temperature $ T_h $ while in contact with a 
high-temperature reservoir.  2) Adiabatic expansion of the gas in thermal 
isolation until the temperature drops to $ T_c $. 3) Isothermal compression 
of the gas while in contact with the low-temperature reservoir. 4) Adiabatic 
compression of the gas until it reaches the temperature $ T_h $. The heat 
engine based on the Carnot cycle is the Carnot engine and its efficiency 
$\eta_{CCE}$ is
\begin{align}
\begin{aligned}
\eta_{CCE} = 1-\frac{T_c}{T_h}.
\end{aligned}
\end{align}
One aims to construct a quantum analogue of the Carnot cycle by replacing the 
classical processes with analogous quantum processes. The two isothermal 
processes in the classical Carnot cycle can be carried over to the quantum 
case since its thermodynamic properties are well-defined. However, it is not 
immediately clear what the quantum analogue of an adiabatic process is. A 
naive thought is to replace classical adiabatic process with quantum adiabatic 
process. The quantum adiabatic process in unitary and thus reversible. But, a 
system which evolves adiabatically reaches a state which is not in thermal 
equilibrium. Defining temperature and other thermodynamic properties for such 
a state may not be possible. Therefore, we need to add an additional step 
which allows the system to reach thermal equilibrium with the reservoir at the 
the end of quantum adiabatic evolution. The proposed quantum Carnot cycle \cite{xiaoConstructionOptimizationQuantum2015} is 
schematically depicted in Fig 1.

\begin{figure}[H]
	\includegraphics[scale=0.3]{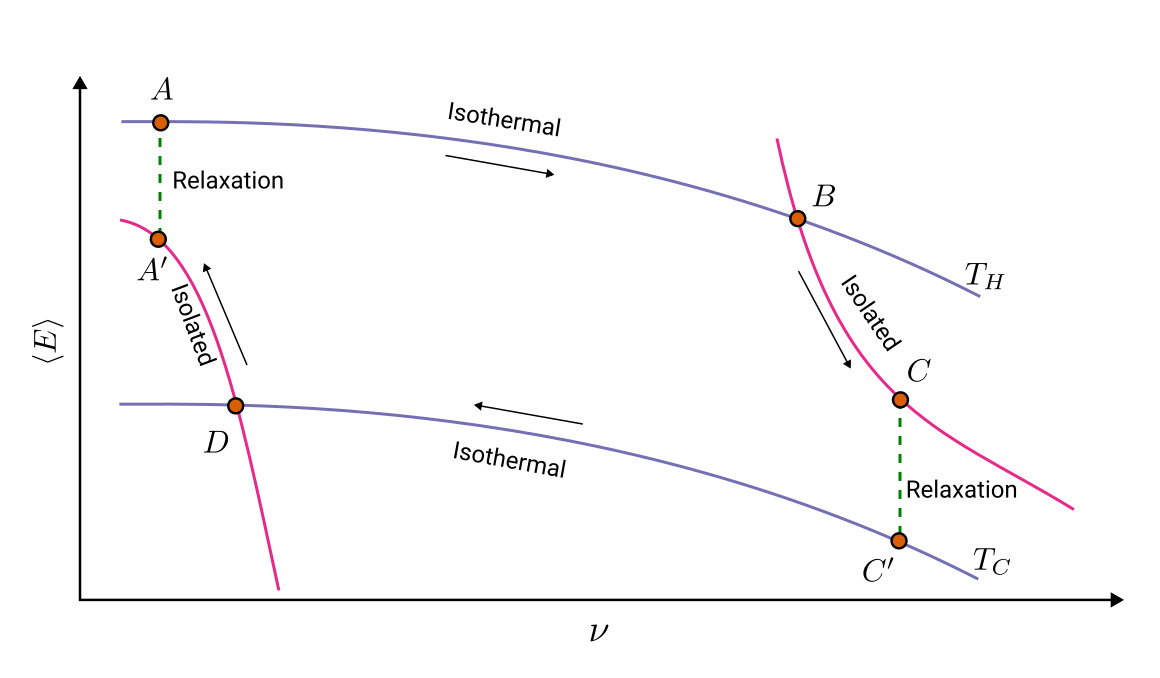}
	\caption{Modified quantum Carnot engine diagram}
\end{figure}
In the quantum Carnot cycle, the processes from $ A \to B $ and $ C \to D $ 
are isothermal processes. The parameter $ \nu $ is assumed to be the only 
system parameter tunable in the cycle operation. The processes from $ B \to C'$
and $ D \to A' $ are quantum adiabatic processes. However, at $ C' $ and $ A'$ 
the system is not in thermal equilibrium. Therefore, at these points we allow 
the system to equilibrate while in contact the reservoir leading to 
processes $ C'\to C $ and $ A'\to A $ by fixing the value of the parameter 
$ \nu $. The total heat dumped into the cold reservoir is 
\begin{align}
\begin{aligned}
Q_{out} &= T_c (S_C-S_D) + \braket{E_{C'} }-\braket{ E_{C}},
\end{aligned}
\end{align}
where $ S_C $ and $ S_D $ describe the entropy of equilibrium states $ C $ and 
$ D $, respectively. Here the term $ \braket{E_{C'} }-\braket{ E_{C}}$ is the
heat dumped into the cold reservoir. Similarly, the total heat that flows into 
the system is 
\begin{align}
\begin{aligned}
Q_{in} &= T_h (S_B-S_A) + \braket{E_{A} }-\braket{ E_{A'}}.
\end{aligned}
\end{align}
Then, the efficiency of the modified quantum Carnot cycle is 
\begin{align}
\begin{aligned}
\eta_{QCE} &=1-\frac{Q_{out}}{Q_{in}} \\
&=1-\frac{T_c (S_C-S_D) + \braket{E_{C'} }-\braket{ E_{C}}}{ T_h (S_B-S_A) + 
\braket{E_{A} }-\braket{ E_{A'}} }
\end{aligned}
\end{align}
We note that the efficiency of quantum Carnot engine $ \eta_{QCE} $ is in 
general lower than the classical Carnot efficiency $ \eta_{CCE} $. This 
is understandable as there is dissipation during equilibration in the 
processes $C\to C^{'}$ and $A^{'} \to A$.   

\section{Anyonic quantum Carnot engine}
In this section, we will discuss the construction of an anyonic quantum 
Carnot engine based on a system of two anyons in a harmonic oscillator 
potential. The system is described by the Hamiltonian
\begin{align}
\begin{aligned}
H = \frac{1}{2m} \rl{\pmb{p}_1^2 +\pmb{p}_2^2}+ \frac{1}{2} m \omega^2 
\rl{\pmb{x}_1^2+\pmb{x}_2^2}.
\end{aligned}
\end{align}
The anyonic wave functions satisfy the condition $\psi\rl{\pmb{x}_2, \pmb{x}_1}
 = e^{i\pi \nu}  \psi\rl{\pmb{x}_1, \pmb{x}_2} $ for an anticlockwise exchange.
In section 2.1, we found the energy eigenstates and computed the energy 
eigenvalues of the Hamiltonian corresponding to two classes (I) and (II). For 
convenience, let us denote the energy eigenstates by 
$ \ket{\psi^{(\nu)}_{j,k,l,m}} $ and corresponding energy eigenvalues by 
$ E^{(\nu)}_{j,k,l,m} $, for a given value of the statistics parameter $\nu $. 

The system has two parameters: one is the oscillator frequency $ \omega $ and 
the other one is the statistics parameter $ \nu $. To construct the quantum 
Carnot cycle, we fix $ \omega $, and choose $ \nu $ to be the system parameter.
Tuning the statistics parameter $ \nu $ means that the quantum statistics of 
particle changes as we change $ \nu $ during the quantum processes.

We start with the system at $ A $ which is in thermal equilibrium with the 
hot reservoir with temperature $ T_h $ and specified by the value of the 
statistics parameter $ \nu_A $.  The state reaches $ B $ isothermally, and 
the statistics parameter changes from $ \nu_A $ to $ \nu_B $. From 
$ B\to C' $, a quantum adiabatic process takes place in thermal isolation 
and the statistics parameter changes from $ \nu_B $ to $ \nu_{C'} $. During 
the relaxation process from $ C' \to C $, the system equilibrates with the 
cold reservoir with temperature $ T_c $ while the statistics parameter remains 
unchanged. From $ C\to D $, the statistics parameter changes from $ \nu_{C'} $ 
to $ \nu_{D} $ and state evolves isothermally while in contact with the cold 
reservoir. From $ D \to D' $ a quantum adiabatic process takes place while 
the statistics parameter gradually changes from $ \nu_{D} \to  \nu_{A} $. 
Finally, the system is allowed to equlibrate with the hot reservoir and 
reach the initial state $ A $ to complete the cycle.

\subsection{Efficiency}
The state of the system at $ A $ is described by the density matrix 
\begin{align}
\begin{aligned}
\rho(A) = \sum_{j,k,l,m} P_{j,k,l,m}(A) \ket{\psi^{(\nu_A)}_{j,k,l,m}}  
\bra{\psi^{(\nu_A)}_{j,k,l,m}},
\end{aligned}
\end{align}
where
\begin{align}
\begin{aligned}
P_{j,k,l,m}(A) = \frac{e^{-\beta_h E^{(\nu_A)}_{j,k,l,m}  }}{Z(A)}.
\end{aligned}
\end{align}
Here $ \beta_h = 1/(k_B T_h)$ is the inverse temperature, and $ Z(A)  $ is the 
partition function $Z(A) = \sum_{j,k,l,m} e^{-\beta_h E^{(\nu_A)}_{j,k,l,m}}$. 
From $A$, the system reaches $B$ which is described by the density matrix
\begin{align}
\begin{aligned}
\rho(B) = \sum_{j,k,l,m} P_{j,k,l,m}(B) \ket{\psi^{(\nu_B)}_{j,k,l,m}}  
\bra{\psi^{(\nu_B)}_{j,k,l,m}},
\end{aligned}
\end{align}
where
\begin{align}
\begin{aligned}
P_{j,k,l,m}(B) = \frac{e^{-\beta_h E^{(\nu_B)}_{j,k,l,m}  }}{Z(B)},  
~~Z(B) = \sum_{j,k,l,m} e^{-\beta_h E^{(\nu_B)}_{j,k,l,m}  }
\end{aligned}
\end{align}
Similarly, one can obtain the density matrices $\rho(C)$ and $\rho(D)$ at $C$ 
and $D$, respectively. 

From these equations, we can compute the efficiency. First, we obtain and 
expression for $Q_{out}$. We have,
\begin{align}
\begin{aligned}
S_C =- k_B\sum_{j,k,l,m} P_{j,k,l,m} (C) \ln  ( P_{j,k,l,m} (C) ).
\end{aligned}
\end{align}
Using the expression for the probability
\begin{align}
\begin{aligned}
P_{j,k,l,m} (C)  = \frac{e^{-\beta_c E^{ (\nu_C)}_{j,k,l,m}}}{Z(C)},
\end{aligned}
\end{align}
we obtain
\begin{align}
\begin{aligned}
S_C &=- k_B\sum_{j,k,l,m} P_{j,k,l,m} (C) \rl{  -\beta_c E^{(\nu_C)}_{j,k,l,m}
 -\ln  ( Z (C) )  } \\
& = \frac{1}{T_c}\braket{E_C} +k_B \ln  ( Z (C) ) .
\end{aligned}
\end{align}
Similarly, we have
\begin{align}
\begin{aligned}
S_D & = \frac{1}{T_c}\braket{E_D} +k_B \ln  ( Z (D) ) .
\end{aligned}
\end{align}
It follows that the total heat dumped into the reservoir is
\begin{align}
\begin{aligned}
Q_{out} &= T_c (S_C-S_D) + \braket{E_{C'} }-\braket{ E_{C}}\\
& = T_c (\frac{1}{T_c}\braket{E_C} +k_B \ln  ( Z (C) )  -\frac{1}{T_c}
\braket{E_D} -k_B \ln  ( Z (D) ) ) + \braket{E_{C'} }-\braket{ E_{C}}\\
& = k_B T_c ln \frac{Z(C)}{Z(D)} - \braket{E_D} +\braket{E_{C'} }
\end{aligned}
\end{align}
Here
\begin{align}
\begin{aligned}
\braket{E_D} & = \frac{1}{Z(D)}\sum_{j,k,l,m} E^{(\nu_D)}_{j,k,l,m}  
e^{-\beta_c   E^{ (\nu_D)}_{j,k,l,m} } \\
\braket{E_{C'}} & = \frac{1}{Z(B)}\sum_{j,k,l,m} E^{ (\nu_C)}_{j,k,l,m} 
e^{-\beta_h   E^{(\nu_B) }_{j,k,l,m} } \\
\end{aligned}
\end{align}
Note that the system is not in a thermal equilibrium at $C'$. But, it is 
possible to compute the expectation value $\braket{E_{C'}}$ since the 
quantum adiabatic processes maintain populations at each quantum level. 

Similarly we can compute the total heat flow into the system:
\begin{align}
\begin{aligned}
Q_{in} & = k_B T_h ln \frac{Z(B)}{Z(A)} +\braket{E_B} -\braket{E_{A'} }
\end{aligned}
\end{align}
where
\begin{align}
\begin{aligned}
\braket{E_B} & = \frac{1}{Z(B)}\sum_{j,k,l,m} E^{ (\nu_B)}_{j,k,l,m} 
e^{-\beta_h   E^{ (\nu_B) }_{j,k,l,m}} \\
\braket{E_{A'}} & = \frac{1}{Z(D)}\sum_{j,k,l,m} E^{ (\nu_A)}_{j,k,l,m} 
e^{-\beta_c   E^{(\nu_D)}_{j,k,l,m}  } \\
\end{aligned}
\end{align}
Therefore, the efficiency is 
\begin{align}
\begin{aligned}
\eta_{QCE} & =1-  \frac{ k_B T_c ln \frac{Z(C)}{Z(D)} - \braket{E_D} 
+\braket{E_{C'} } }{ k_B T_h ln \frac{Z(B)}{Z(A)} + \braket{E_B} 
-\braket{E_{A'} }  }
\end{aligned}
\end{align}
The efficiency can be computed exactly by finding closed form expressions for 
the infinite sums involved in the above expression. The details are relegated 
to the appendix. The efficiency is
\begin{align}
\begin{aligned}
\eta_{QCE} & =1-  \frac{ k_B T_c ln \frac{\mathcal{Z}(\nu_{C'}, \beta_c)}
{\mathcal{Z}(\nu_{D}, \beta_c)} - \mathcal{E}(\nu_D,\nu_D,\beta_c) 
+\mathcal{E}(\nu_C,\nu_B,\beta_h)  }{ k_B T_h ln \frac{\mathcal{Z}(\nu_{B}, 
\beta_h)}{\mathcal{Z}(\nu_{A}, \beta_h)} - \mathcal{E}(\nu_B,\nu_B,\beta_h) 
+\mathcal{E}(\nu_A,\nu_D,\beta_c)},
\end{aligned}
\end{align}
where
\begin{align}
\begin{aligned}
\mathcal{Z}(\nu,\beta) & = \frac{\cosh ((1-\nu) \beta \hbar \omega )}
{8\sinh^2 \rl{\frac{\beta \hbar \omega }{2}}\sinh^2 \rl{\beta \hbar \omega} },
\end{aligned}
\end{align}
and
\begin{align}
\begin{aligned}
&\mathcal{E}(\nu,\nu',\beta) =
\frac{1}{2 \sinh (\beta 
	\hbar \omega)  \cosh\left(\beta \hbar \omega(1-\nu')\right)  }\\ 
&\times\left((\nu+2) \cosh \left(\beta \hbar \omega  
\left(\nu'-2\right)\right)-(\nu -4) \cosh \left(\beta  \hbar \omega  
\nu '\right)+2 \cosh \left(\beta  \hbar \omega (1- \nu')\right)\right)
\end{aligned}
\end{align}
\section{Conclusion}
This paper is a sequel to \cite {maniQuantumThermodynamicsSmall2024} in which 
the quantum Otto engine of a small number of one-dimensional anyons was 
studied. Here we extend the results to two-dimensional anyons in a harmonic 
trap which are used as the working medium for a quantum Carnot engine. A 
general formula is derived for the efficiency of the anyonic Carnot engine. 
It is found that the efficiency of the quantum Carnot engine is less than the 
efficiency of the classical Carnot engine. Nevertheless, the promising feature 
associated with extracting energy from purely quantum sources which can be 
converted into mechanical work in quantum thermodynamic engines makes the 
study of such systems interesting. 

It is obvious that one can also construct quantum thermodynamic engines 
based on Exclusion Statistics \cite{myersThermodynamicsStatisticalAnyons2021}, and also non-abelian anyons. We hope to 
report on these cases in the near future.  

\section{Acknowledgement:} This work is partially supported by a grant to 
CMI from the Infosys Foundation. 

%\bibliographystyle{unsrt}
%\bibliography{aqce.bib}

\begin{thebibliography}{10}

\bibitem{maniQuantumThermodynamicsSmall2024}
H.~S. Mani, N.~Ramadas, and V.~V. Sreedhar.
\newblock Quantum thermodynamics of small systems: {{The}} anyonic otto engine.
\newblock {\em Modern Physics Letters A}, 39(08):2450020, March 2024.

\bibitem{binderThermodynamicsQuantumRegime2018}
Felix Binder, Luis~A. Correa, Christian Gogolin, Janet Anders, and Gerardo
  Adesso, editors.
\newblock {\em Thermodynamics in the {{Quantum Regime}}: {{Fundamental
  Aspects}} and {{New Directions}}}, volume 195 of {\em Fundamental
  {{Theories}} of {{Physics}}}.
\newblock Springer International Publishing, Cham, 2018.

\bibitem{myersQuantumThermodynamicDevices2022}
Nathan~M. Myers, Obinna Abah, and Sebastian Deffner.
\newblock Quantum thermodynamic devices: {{From}} theoretical proposals to
  experimental reality.
\newblock {\em AVS Quantum Science}, 4(2):027101, April 2022.

\bibitem{kochQuantumEngineBEC2023}
Jennifer Koch, Keerthy Menon, Eloisa Cuestas, Sian Barbosa, Eric Lutz,
  Thom{\'a}s Fogarty, Thomas Busch, and Artur Widera.
\newblock A quantum engine in the {{BEC}}--{{BCS}} crossover.
\newblock {\em Nature}, 621(7980):723--727, September 2023.

\bibitem{feshbachSmallSystemsWhen1988}
H.~Feshbach.
\newblock Small systems: When does thermodynamics apply?
\newblock {\em IEEE Journal of Quantum Electronics}, 24(7):1320--1322, July
  1988.

\bibitem{hillThermodynamicsSmallSystems2013}
Terrell~L. Hill.
\newblock {\em Thermodynamics of {{Small Systems}}, {{Parts I}} and {{II}}}.
\newblock Dover {{Books}} on {{Chemistry Series}}. Dover Publications,
  Incorporated, Newburyport, 1st ed edition, 2013.

\bibitem{callenThermodynamicsIntroductionThermostatistics1985}
Herbert~B. Callen.
\newblock {\em Thermodynamics and an Introduction to Thermostatistics}.
\newblock Wiley, New York, 2nd ed edition, 1985.

\bibitem{leggettQuantumLiquidsBose2006}
Anthony~J. Leggett.
\newblock {\em Quantum Liquids: {{Bose}} Condensation and {{Cooper}} Pairing in
  Condensed-Matter Systems}.
\newblock Oxford Graduate Texts. Oxford University Press, Oxford New York,
  2006.

\bibitem{leinaasTheoryIdenticalParticles1977}
J.~M. Leinaas and J.~Myrheim.
\newblock On the theory of identical particles.
\newblock {\em Il Nuovo Cimento B (1971-1996)}, 37(1):1--23, January 1977.

\bibitem{myrheimAnyons1999}
J.~Myrheim.
\newblock Anyons.
\newblock In A.~Comtet, T.~Jolic{\oe}ur, S.~Ouvry, and F.~David, editors, {\em
  Aspects Topologiques de La Physique En Basse Dimension. {{Topological}}
  Aspects of Low Dimensional Systems}, pages 265--413, Berlin, Heidelberg,
  1999. Springer.

\bibitem{xiaoConstructionOptimizationQuantum2015}
Gaoyang Xiao and Jiangbin Gong.
\newblock Construction and optimization of a quantum analog of the {{Carnot}}
  cycle.
\newblock {\em Physical Review E}, 92(1):012118, July 2015.

\bibitem{myersThermodynamicsStatisticalAnyons2021}
Nathan~M. Myers and Sebastian Deffner.
\newblock Thermodynamics of {{Statistical Anyons}}.
\newblock {\em PRX Quantum}, 2(4):040312, October 2021.


\end{thebibliography}

\appendix
\section{Calculation of the efficiency}
We note the closed form expressions for the following infinite sums
\begin{align}
\begin{aligned}
\sum_{m=0}^\infty e^{-\gamma m} = \frac{1}{1-e^{-\gamma}} = 
\frac{e^{\frac{\gamma}{2}}}{2 \sinh \rl{ \frac{\gamma}{2}}}\\
\sum_{m=0}^\infty m e^{-\gamma m} = \frac{1}{4 \sinh^2 \rl{\frac{\gamma}{2}}}\\
\end{aligned}
\end{align}
Using these, we have
\begin{align}
\begin{aligned}
\mathcal{Z}(\nu,\beta) &= \sum_{j,k,l,m=0}^{\infty} 
e^{-\beta E_{j,k,l,m}(\nu)} \\
& = \sum_{j,k,l,m=0}^{\infty} e^{-\beta  \rl{2+\nu+j+k+2l+2m} \hbar \omega} 
+ \sum_{j,k,l,m=0}^{\infty} e^{-\beta \rl{4-\nu+j+k+2l+2m} \hbar \omega } \\
& = \frac{e^{-\beta \hbar \omega  \rl{2+\nu}} +e^{-\beta \hbar \omega 
\rl{4-\nu}}}{ \rl{1-e^{-\beta \hbar \omega } }^2  \rl{1-e^{-2\beta \hbar 
\omega } }^2} \\
& = \frac{\cosh ((1-\nu) \beta \hbar \omega )}{ 32 \sinh^4 \rl{\frac{\beta 
\hbar \omega }{2}}    \cosh^2 \rl{\frac{\beta \hbar \omega }{2}}   }
\end{aligned}
\end{align}
We have $ Z(A) = \mathcal{Z}(\nu_A,\beta_h) ,Z(B) = \mathcal{Z}(\nu_B,\beta_h),
Z(C) = \mathcal{Z}(\nu_C,\beta_c) ,Z(D) = \mathcal{Z}(\nu_D,\beta_c) $.
Also, we would like to evaluate the sum
\begin{align*}
&\mathcal{E}(\nu,\nu',\beta)  \\
=& \frac{1}{Z(\nu',\beta) }\sum_{j,k,l,m} E_{j,k,l,m}(\nu)  
e^{-\beta E_{j,k,l,m} (\nu')} \\
=& \frac{1}{Z(\nu',\beta) } \bigg(  \sum_{j,k,l,m=0}^{\infty} 
\rl{2+\nu+j+k+2l+2m}\hbar \omega e^{-\beta  \rl{2+\nu'+j+k+2l+2m}\hbar\omega}
\\&+ \sum_{j,k,l,m=0}^{\infty} \rl{4-\nu+j+k+2l+2m} \hbar \omega 
e^{-\beta \rl{4-\nu'+j+k+2l+2m} \hbar \omega } \bigg) \\
%----------------
=& \frac{1}{Z(\nu',\beta) } \bigg(  \sum_{j,k,l,m=0}^{\infty} 
\rl{2+\nu+j+k+2l+2m}\hbar \omega e^{-\beta  \rl{2+\nu'+j+k+2l+2m} 
\hbar \omega} \\&+ \sum_{j,k,l,m=0}^{\infty} \rl{4-\nu+j+k+2l+2m} 
\hbar \omega e^{-\beta \rl{4-\nu'+j+k+2l+2m} \hbar \omega } \bigg) \\
%----------------
& =  \frac{\hbar \omega e^{-\beta  \rl{2+\nu'}\hbar \omega} }{Z(\nu',\beta) } 
\bigg( (2+\nu) \frac{e^{3 \beta \hbar \omega }}{32 \sinh^4\rl{\frac{\beta\hbar 
\omega }{2}} \cosh^2 \rl{\frac{\beta \hbar \omega }{2}}  }  + 
\frac{ 3 e^{2 \beta \hbar \omega } +e^{3\beta \hbar \omega }  }{128  \sinh^5 
\rl{\frac{\beta \hbar \omega }{2}} \cosh^3 \rl{\frac{\beta \hbar \omega }{2}}}
\bigg) \\
& + \frac{\hbar \omega e^{-\beta  \rl{4-\nu'} \hbar \omega} }{Z(\nu',\beta) }
\bigg( (4-\nu)\frac{e^{3 \beta \hbar \omega }}{ 32 \sinh^4 \rl{\frac{\beta 
\hbar \omega }{2}}    \cosh^2 \rl{\frac{\beta \hbar \omega }{2}}  }  + 
\frac{ 3 e^{2 \beta \hbar \omega } +e^{3\beta \hbar \omega }  }
{128 \sinh^5 \rl{\frac{\beta \hbar \omega }{2}}\cosh^3 \rl{\frac{\beta 
\hbar \omega }{2}}  }    \bigg)\\
%--------------
%----------------
& =  \frac{\hbar \omega e^{-\beta  \rl{2+\nu'}\hbar \omega} }{\cosh ((1-\nu')
\beta \hbar \omega ) } \bigg( (2+\nu) e^{3 \beta \hbar \omega }  + 
\frac{ 3 e^{2 \beta \hbar \omega } +e^{3\beta \hbar \omega }  }
{4\sinh \rl{\frac{\beta \hbar \omega }{2}} \cosh \rl{\frac{\beta\hbar\omega }
{2}}  }     \bigg) \\
& + \frac{\hbar \omega e^{-\beta  \rl{4-\nu'} \hbar \omega} }{\cosh ((1-\nu') 
\beta \hbar \omega )}  \bigg( (4-\nu)e^{3 \beta \hbar \omega }  + 
\frac{ 3 e^{2 \beta \hbar \omega } +e^{3\beta \hbar \omega }  }
{4\sinh\rl{\frac{\beta\hbar \omega }{2}}\cosh \rl{\frac{\beta\hbar\omega }{2}}
}     \bigg)\\
%--------------
%----------------
&=\frac{1}{2 \sinh (\beta 
	\hbar \omega)  \cosh\left(\beta \hbar \omega(1-\nu')\right)  } 
\left((\nu+2) \cosh \left(\beta \hbar \omega  \left(\nu'-2\right)\right)
-(\nu -4) \cosh \left(\beta\hbar\omega\nu '\right)+2 \cosh\left(\beta\hbar
 \omega (1- \nu')\right)\right)
%-------------
\end{align*}
Using the above expression, we can compute $ \braket{E_D} = \mathcal{E}(\nu_D,
\nu_D,\beta_c),\braket{E_{C'}} = \mathcal{E}(\nu_C,\nu_B,\beta_h),
\braket{E_{B}} = \mathcal{E}(\nu_B,\nu_B,\beta_h),\braket{E_{A'}} = 
\mathcal{E}(\nu_A,\nu_D,\beta_c) $.
Therefore, the efficiency
\begin{align}
\begin{aligned}
\eta_{QCE} & =1-  \frac{ k_B T_c ln \frac{\mathcal{Z}(\nu_{C'}, \beta_c)}
{\mathcal{Z}(\nu_{D}, \beta_c)} - \mathcal{E}(\nu_D,\nu_D,\beta_c) 
+\mathcal{E}(\nu_C,\nu_B,\beta_h)}{ k_B T_h ln \frac{\mathcal{Z}(\nu_{B}, 
\beta_h)}{\mathcal{Z}(\nu_{A}, \beta_h)} - \mathcal{E}(\nu_B,\nu_B,\beta_h) 
+\mathcal{E}(\nu_A,\nu_D,\beta_c)},
\end{aligned}
\end{align}

\end{document}